\documentclass[12pt]{iopart}
\usepackage{multicol}
\usepackage[left=3.0cm]{geometry}
\usepackage[utf8]{inputenc}
\usepackage{graphicx}
\usepackage{alltt}
\usepackage{url}
\usepackage{textcomp}
\usepackage{iopams}
\usepackage{multirow}
\usepackage{dcolumn}
\usepackage{booktabs}
\usepackage[multiple]{footmisc}
\usepackage[T1]{fontenc}
\begin{document}

\title[Effects of the circularly polarized beam of linearized gravitational waves]{Effects of the circularly polarized beam of linearized gravitational waves}

\author{W Barker$^\dag$}

\address{$^\dag$ Cavendish Laboratory, Cambridge University, JJ Thomson Avenue, Cambridge, CB3 0HE, UK}

\begin{abstract}
Solutions of the linearized Einstein equations are found that describe a transversely confined beam of circularly polarized gravitational waves on a Minkowski backdrop. By evaluating the cycle-averaged stress-energy-momentum pseudotensor of Landau \& Lifshitz it is found that the angular momentum density is concentrated in the `skin' at the edge of the beam where the intensity falls, and that the ratio of angular momentum to energy per unit length of the beam is $2/\omega$, where $\omega$ is the wave frequency, as expected for a beam of spin-$2$ gravitons. For sharply-defined, uniform, axisymmetric beams, the induced background metric is shown to produce the gravomagnetic field and frame-dragging effects of a gravitational solenoid, whilst the angular momentum current helically twists the space at infinite radius along the beam axis.
\end{abstract}

\\
\\
\submitto{arXiv (preprint for \CQG)}

\section{Introduction}
\paragraph{}
The 1975 edition of Classical Electrodynamics by Jackson\textsuperscript{\cite{jackson1}} poses the problem of finding the electromagnetic field of a circularly polarized beam of light. The suggested solution satisfies the vanishing divergence of the electric field but not the wave equation, so that the intensity of the beam should vary only very slowly across its width compared to the wavelength. The gravitational effects of such a beam (or something close to it) are invstigated by Bonnor\textsuperscript{\cite{nullfluid}}, who found exact solutions to the Einstein equations for beams of \textit{null dust}, grains of which fall along null geodesics without the considerations of finite wavelength (c.f. photons) and which, for the representation of circularly polarized waves may be chosen to spin. Most recently, efforts have been made by Lynden-Bell and Bi\v{c}\'{a}k to complete the picture by reconciling Bonnor's solutions with full classical electromagnetic sources. Since the angular momentum density of Jackson's solution is concentrated in the `skin' of an otherwise uniform beam, the problem lends itself to a study of gravomagnetic effects: it is well known\textsuperscript{\cite{cylinders}}\textsuperscript{\cite{cylinders2}} that such gravitational solenoids, both of material and gravitational sources enclose a uniform gravomagnetic field which rotates the inertial frames within. Furthermore\textsuperscript{\cite{cylinders2}}, an infinite observer of such a material cylindrical shell which is simultaneously transmitting a torque along its length is known to additionally find themselves in a apace wrapped helically around the cylinder axis.
\par
An escalation of this problem is the equivalent beam of circularly polarized gravitational waves. It is the simplest notion of gravitational waves, that emergent from \textit{linearized gravitation}, that bears the closest resemblance to its electromagnetic counterpart, so in this regime the problem is fairly well posed: we do \textit{not} attempt an exact solution of the Einstein equations. In order to relate the `metric' produced by the beam to the beam itself, which is of course only a perturbation of the very same metric, we adapt the approach first formulated by Isaacson\textsuperscript{\cite{average}} and later expounded by Lifshitz\textsuperscript{\cite{vol2}}: a wave-like metric perturbation may `massage' the spacetime across which it ripples into a further non-oscillating `background' perturbation, and the interesting physics is contained therein.
\par
\Sref{1} is devoted to the linearised gravitational analogy of Jackson's solution. Using the energy momentum pseudotensor of Landau and Lifshitz it is shown that the resultant description is consistent with that of a beam of gravitons, each bearing angular momentum $2\hbar$ and energy $\hbar\omega$; this is similar to the null dust picture, and the analogous result for photons is also mentioned in \cite{jackson1}. \Sref{2} is a search for the background metric (with a short reminder of the established method) and a discussion of its place in the theory. Latin indices run over four dimensions and Greek indices run over spatial dimensions only.
\section{Linearized gravitation and the graviton spin-energy ratio} \label{1}
\paragraph{}
We consider a beam of linearized gravitational waves propagating down the $z$-axis. As is usual, we write this as a perturbation $h_{jk}$ with $h=h^{k}_{k}$ of the underlying Minkowski metric, $\eta_{jk}$,
\begin{eqnarray}
g_{jk}=\eta_{jk}+h_{jk}
\end{eqnarray}
and enforce the gravitational Lorentz gauge condition,
\begin{eqnarray} \label{eq:lorentz}
\left(h^{j}_{k}-\frac{1}{2}\delta^{j}_{k}h\right)_{,j}=0,
\end{eqnarray}
from which the linearized Einstein equations become wave-like:
\begin{eqnarray} \label{eq:waves}
h^{i,k}_{j\ ,k}=0.
\end{eqnarray}
Within the gravitational Lorentz gauge the usual coordinate transformations, $x'^{i}=x^{i}+\xi^{i}$, where $\xi^{i,k}_{\ \ ,k}=0$, are made to move to the \textit{transverse traceless} (TT) gauge: $h^{0}_{k}=h=0$. From the Fourier expansion,
\begin{eqnarray}
h^{j}_{m}=\int\mathrm{d}^{3}k\left( a^{j}_{m}e^{-ik_{l}x^{l}}+{a^*}^{j}_{m}e^{ik_{l}x^{l}}\right),
\end{eqnarray}
where $a^{j}_{m}=a^{j}_{m}\left(x, y\right)$ and $k^{i}=\left(\omega/c, 0, 0, k\right)$ and since the real and imaginary parts of $a^{j}_{k}e^{ik_{l}x^{l}}$ independently satisfy \Eref{eq:lorentz} in the TT gauge,
\begin{eqnarray}
a^{l}_{j,l}-ik_{l}a^{l}_{j}=\partial_{x}a_{1\beta}+	\partial_{y}a_{2\beta}+ika_{3\beta}=0,
\end{eqnarray}
which, writing $\varphi=-k^{l}x_{l}$, gives a general solution of the Lorentz gauge constraint presented in \Tref{tab2}.
\par
It remains to satisfy the d'Alembertian of \Eref{eq:waves}, which is trivial for the case of plane waves where the $a^{j}_{k}$ are constant. If however the Fourier coefficients vary on a wavelength $K=\Omega/c$ such that $\Omega\ll\omega$, operators of the form $k^{-2}\partial^{2}_{ab}$ where $a$, $b$ are $1$ or $2$ could be ignored and so therefore could $h_{33}$. Then, enforcing tracelessness there is the usual choice of plus and cross-polarized gravity waves: $a_{11}=-a_{22}$, $a_{12}=a_{21}$. For the circularly polarized case we define some sufficiently slowly varying transverse envelope $f\left(x, y\right)$ such that $a_{11}=f\left(x, y\right)/2$, $a_{12}=-if\left(x, y\right)/2$. This yields to $\mathcal{O}\left(K^{2}f\right)$,
\begin{table}
	\caption{\label{tab2}The components of $h_{jk}$ up to solution of the Lorentz gauge.}
	\begin{tabular}[h]{@{}|c|l|}
		\hline
		$h_{11}$ & $a_{11}e^{i\varphi}+{a^*}_{11}e^{-i\varphi}$ \\ 
		$h_{12}$ & $a_{12}e^{i\varphi}+{a^*}_{12}e^{-i\varphi}$ \\ 
		$h_{22}$ & $a_{22}e^{i\varphi}+{a^*}_{22}e^{-i\varphi}$ \\ 
		$h_{13}$ & $\frac{i}{k}\left[\left(\partial_{x}a_{11}+	\partial_{y}a_{12}\right)e^{i\varphi}-\left(\partial_{x}{a^*}_{11}+	\partial_{y}{a^*}_{12}\right)e^{-i\varphi}\right]$ \\ 
		$h_{23}$ & $\frac{i}{k}\left[\left(\partial_{x}a_{21}+	\partial_{y}a_{22}\right)e^{i\varphi}-\left(\partial_{x}{a^*}_{21}+	\partial_{y}{a^*}_{22}\right)e^{-i\varphi}\right]$ \\ 
		$h_{33}$ & $-\frac{1}{k^{2}}\left[ \left(\partial^{2}_{xx}a_{11}+2\partial^{2}_{xy}a_{12}+\partial^{2}_{yy}a_{22}\right)e^{i\varphi}+\left(\partial^{2}_{xx}{a^*}_{11}+2\partial^2_{xy}{a^*}_{12}+\partial^2_{yy}{a^*}_{22}\right)e^{-i\varphi}\right]$ \\ 
		\hline
	\end{tabular}
\end{table}
\begin{eqnarray} \label{eq:perturbation}
-\left[h_{\alpha\beta}\right]= \left( 
\begin{array}{ccc}
f\cos\varphi & f\sin\varphi & \frac{1}{k}\left(\cos\varphi\partial_{y}f-\sin\varphi\partial_{x}f\right) \\
\cdot\cdot\cdot & -f\cos\varphi & \frac{1}{k}\left(\cos\varphi\partial_{x}f+\sin\varphi\partial_{y}f\right) \\
\cdot\cdot\cdot & \cdot\cdot\cdot & 0 \\
\end{array} 
\right),
\end{eqnarray}
so that it could generally be said that $\mathcal{O}\left(h^{i}_{k}\right)=\mathcal{O}\left(f\right)$, and $f\left(x, y\right)$ is such that its $n$th transverse derivatives are $\mathcal{O}\left(c^{-n}\Omega^{n}f\right)$. In the TT gauge, plane gravitational waves appear as perturbations of those elements of the metric with any indices other than those in which $k^{i}$ lies. In \Eref{eq:perturbation} the plane wave is transversely confined by adding `perpendicular' components, a principle that reflects the electromagnetic solution found in \cite{jackson1}:
\begin{eqnarray} \label{jackson}
\mathbf{E}=\left(f\cos\varphi, -f\sin\varphi, -\frac{1}{k}\left(\cos\varphi\partial_{y}f+\sin\varphi\partial_{x}f\right)\right),
\end{eqnarray}
where $f\left(x, y\right)$ is similarly defined and the vector $\mathbf{B}$ has the same form. The analogy is not exact: \Eref{eq:perturbation} is a wave in what amount to field potentials (the metric rather than the Christoffel symbols) whereas \Eref{jackson} refers to the fields themselves.
\paragraph{}
In linearized gravitation, the energy-momentum pseudotensor of Landau and Lifshitz is
\begin{eqnarray}
t^{ik}=\frac{c^{4}}{16\pi G}\left(h^{ik}_{\ \ ,l}h^{lm}_{\ \ ,m}-h^{il}_{\ \ ,l}h^{km}_{\ \ ,m}+\frac{1}{2}\eta^{ik}h^{ln}_{\ \ ,p}h^{p}_{l,n}-h^{kn}_{\ \ ,p}h^{p,i}_{n}\right.\nonumber \\ \left.-h^{in}_{\ \ ,p}h^{p,k}_{n}+h^{il}_{\ \ ,n}h^{k,n}_{l}+\frac{1}{2}h^{n,i}_{q}h^{q,k}_{n}-\frac{1}{4}h^{n,i}_{n}h^{p,k}_{p}-\frac{1}{4}\eta^{ik}h^{n}_{q,l}h^{q,l}_{n}\right.\nonumber \\ \left.+\frac{1}{8}\eta^{ik}h^{n}_{n,l}h^{p,l}_{p}\right),
\end{eqnarray}
however the traceless Lorentz condition removes the first, second, eighth and tenth terms. For pure plane waves down $z$ in the TT gauge only derivatives in indices $0$ and $3$ remain, whilst $h_{\alpha\beta}$ vanishes except for combinations of the indices $1$ and $2$. Therefore the third, fourth and fifth terms also vanish. Finally the sixth, ninth (and tenth) terms are of the form $T^{...}_{...,p}T^{...,p}_{...}$, and may be re-written with a vanishing factor of $k^{p}k_{p}$. Thus for pure plane waves, there remains only the well-known simple form,
\begin{eqnarray}
t^{ik}=\frac{c^{4}}{32\pi G}h^{n,i}_{q}h^{q,k}_{n}.
\end{eqnarray}
In the present case however, it is no longer possible to make the final two simplifications and the appropriate pseudotensor is
\begin{eqnarray} \label{pseudotensor_remainder}
t^{ik}=\frac{c^{4}}{16\pi G}\left[\frac{1}{2}\eta^{ik}\left(h^{ln}_{\ \ ,p}h^{p}_{l,n}-\frac{1}{2}h^{n}_{q,l}h^{q,l}_{n}\right)-h^{kn}_{\ \ ,p}h^{p,i}_{n}-h^{in}_{\ \ ,p}h^{p,k}_{n}\right.\nonumber \\ \left.+h^{il}_{\ \ ,n}h^{k,n}_{l}+\frac{1}{2}h^{n,i}_{q}h^{q,k}_{n}\right].
\end{eqnarray}
By taking the cycle-average, denoted by $\left\langle\right\rangle$, of leading-order terms of relevant elements of \Eref{pseudotensor_remainder}, it is found that
\begin{eqnarray}
\left\langle t^{00}\right\rangle=\frac{c^{2}\omega^{2}}{16\pi G}f^{2}, \quad \left\langle t^{01}\right\rangle=-\frac{c^{3}\omega}{8\pi G}f\partial_{y}f, \quad \left\langle t^{02}\right\rangle=\frac{c^{3}\omega}{8\pi G}f\partial_{x}f,
\end{eqnarray}
from which the density of the component of angular momentum that lies along the beam and the energy density are
\begin{eqnarray} \label{eq:dens}
\left\langle \lambda_{z}\right\rangle=\frac{c^{2}\omega}{8\pi G}\left(xf\partial_{x}f+yf\partial_{y}f\right), \quad \left\langle \mathcal{E}\right\rangle=\frac{c^{2}\omega^{2}}{16\pi G}f^{2}.
\end{eqnarray}
Integrating by parts over the whole plane perpendicular to the beam gives
\begin{eqnarray}
\frac{\int\int\left\langle \lambda_{z}\right\rangle\mathrm{d}x\mathrm{d}y}{\int\int\left\langle \mathcal{E}\right\rangle\mathrm{d}x\mathrm{d}y}=\frac{2}{\omega},
\end{eqnarray}
which agrees with the expectation that gravitons have energy $\hbar\omega$ and spin $2\hbar$. An analogous result for photons is mentioned in \cite{jackson1}.
\section{The background metric} \label{2}
\paragraph{}
If the background metric were not Minkowski, the gravitational Lorentz condition and wave equation would take the form:
\begin{eqnarray} \label{eq:fullwavelorentz}
h^{k}_{i;k}=h^{\ \ ;l}_{ik\ ;l}=0.
\end{eqnarray}
As is developed in \cite{vol2}, based on the results of \cite{average} the full metric may be written
\begin{eqnarray} \label{eq:fullmetric}
g_{jk}=\eta_{jk}+s_{jk}+h_{jk},
\end{eqnarray}
where $s_{jk}$ is some distortion of the background metric which is, by default, zero-order in the $h_{jk}$ (by which in particular it is \textit{non-oscillating}) and which is in fact `induced' by its presence (no other sources being considered here). This comes about because of the exact form of the Ricci tensor,
\begin{eqnarray} \label{eq:ricci}
R_{jk}=\Gamma^{l}_{ik,l}-\Gamma^{l}_{il,k}+\Gamma^{l}_{ik}\Gamma^{m}_{lm}-\Gamma^{m}_{il}\Gamma^{l}_{km},
\end{eqnarray}
which may be evaluated for \Eref{eq:fullmetric} to give terms to up to second order in the wave-like $h_{jk}$: $R_{jk}=R^{\left(0\right)}_{\ \ \ jk}+R^{\left(1\right)}_{\ \ \ jk}+R^{\left(2\right)}_{\ \ \ jk}+\mathcal{O}\left(h^{3}\right)$; the same may also be said of the exact $t_{jk}$. Since terms to first order in the $h_{jk}$ vanish upon averaging, the averaged exact Einstein equations take the form $R^{\left(0\right)}_{\ \ \ jk}=-\left\langle R^{\left(2\right)}_{\ \ \ jk}\right\rangle$. The $t^{ik}$ is constructed to satisfy
\begin{eqnarray} \label{eq:pseudotensormotivation}
\left(-g\right)\left[\frac{c^{4}}{8\pi G}\left(R^{ik}-\frac{1}{2}g^{ik}R\right)+t^{ik}\right]=\zeta^{ikl}_{\ \ \ ,l},
\end{eqnarray}
where the (non-tensorial) $\zeta^{ikl}$ is skew-symmetric in its last pair of indices. Taking the average of this equation one can retain on the RHS only those largest terms of $\mathcal{O}\left(\omega^{2}f^{2}\right)$ (on dimensional grounds factors of $\omega$ replace those of $\Omega$ and vice-versa). Since the RHS was written as a four-divergence, it may be evaluated using Gauss' law, causing it to drop an order in $\omega$. If the equation is satisfied by pairwise cancellation on both sides of terms arising from various orders of the small $f$ and then large $\omega$ or small $\Omega$, we are left to $\mathcal{O}\left(\omega^{2}f^{2}\right)$ with
\begin{eqnarray} \label{eq:kernel}
R^{(0)}_{\ \ \ ik}-\frac{1}{2}\eta_{ik}R^{(0)}=\frac{8\pi G}{c^{4}}\left\langle t^{(2)}_{\ \ \ ik} \right\rangle
\end{eqnarray}
and we may recover $s_{jk}$ by the simultaneous solution of \Eref{eq:kernel} with \Eref{eq:fullwavelorentz}: this is the extent of the method described in \cite{vol2}.
\begin{table}
\caption{\label{tab1}The components of \Eref{eq:kernel}, good to the scale in the first column.}
\begin{tabular}[h]{@{}|c|c|l|}
\hline
Order & Indices $ik$ & Relation \\
\hline
\multirow{6}{*}{$\mathcal{O}\left(c^{-2}\omega^{2}f^{2}\right)$}& $00$ & $\frac{1}{2}\left(\nabla^{2}_{\perp}s^{(0)}_{\ \ \ 33}+\partial^{2}_{xx}s^{(0)}_{\ \ \ 22}-2\partial^{2}_{xy}s^{(0)}_{\ \ \ 12}+\partial^{2}_{yy}s^{(0)}_{\ \ \ 11}\right)=\frac{\omega^{2}f^{2}}{2c^{2}}$ \\ 
& $03$ & $-\frac{1}{2}\nabla^{2}_{\perp}s^{(0)}_{\ \ \ 03}=\frac{\omega^{2}f^{2}}{2c^{2}}$ \\
& $33$ & $\frac{1}{2}\left(\nabla^{2}_{\perp}s^{(0)}_{\ \ \ 00}-\partial^{2}_{xx}s^{(0)}_{\ \ \ 22}+2\partial^{2}_{xy}s^{(0)}_{\ \ \ 12}-\partial^{2}_{yy}s^{(0)}_{\ \ \ 11}\right)=\frac{\omega^{2}f^{2}}{2c^{2}}$ \\
& $11$ & $\frac{1}{2}\partial^{2}_{yy}\left(s^{(0)}_{\ \ \ 00}-s^{(0)}_{\ \ \ 33}\right)=0$ \\
& $12$ & $-\frac{1}{2}\partial^{2}_{xy}\left(s^{(0)}_{\ \ \ 00}-s^{(0)}_{\ \ \ 33}\right)=0$ \\
& $22$ & $\frac{1}{2}\partial^{2}_{xx}\left(s^{(0)}_{\ \ \ 00}-s^{(0)}_{\ \ \ 33}\right)=0$ \\
\hline
\multirow{4}{*}{$\mathcal{O}\left(c^{-2}\omega\Omega f^{2}\right)$}& $01$ & $-\frac{1}{2}\left(\partial^{2}_{yy}s^{(1)}_{\ \ \ 01}-\partial^{2}_{xy}s^{(1)}_{\ \ \ 02}\right)=-\frac{\omega f\partial_{y}f}{c}$ \\
& $02$ & $-\frac{1}{2}\left(\partial^{2}_{xx}s^{(1)}_{\ \ \ 02}-\partial^{2}_{xy}s^{(1)}_{\ \ \ 01}\right)=\frac{\omega f\partial_{x}f}{c}$ \\
& $31$ & $\frac{1}{2}\left(\partial^{2}_{yy}s^{(1)}_{\ \ \ 13}-\partial^{2}_{xy}s^{(1)}_{\ \ \ 23}\right)=-\frac{\omega f\partial_{y}f}{c}$ \\
& $32$ & $\frac{1}{2}\left(\partial^{2}_{xx}s^{(1)}_{\ \ \ 23}-\partial^{2}_{xy}s^{(1)}_{\ \ \ 13}\right)=\frac{\omega f\partial_{x}f}{c}$ \\ 
\hline
\end{tabular}
\end{table}
\paragraph{}
In order to gain traction it is a great simplification to consider the regime where $s_{jk}$ is small, specifically small enough that we may neglect its Christoffel symbols in \Eref{eq:fullwavelorentz}. We expect averaged quantities to vary only transversely to the beam and to do so with characteristic frequency $\Omega$, as does $f$. Furthermore, we will neglect the latter two terms in \Eref{eq:ricci}, and by doing so we are effectively linearizing the theory of this background field. If $s_{jk}$ were expanded in powers of $\Omega/\omega$, the largest components of \Eref{eq:kernel}, relevant to $\mathcal{O}\left(c^{-2}\omega^{2}f^{2}\right)$, (see \Tref{tab1}) give
\begin{eqnarray}
\nabla^{2}_{\perp}s^{(0)}_{\ \ \ 00}=\nabla^{2}_{\perp}s^{(0)}_{\ \ \ 33}=-\nabla^{2}_{\perp}s^{(0)}_{\ \ \ 03}=\omega^{2}f^{2}/\left(c^{2}\right),
\end{eqnarray}
whilst
\begin{eqnarray} \label{second}
\partial^{2}_{xx}s^{(0)}_{\ \ \ 22}-2\partial^{2}_{xy}s^{(0)}_{\ \ \ 12}+\partial^{2}_{yy}s^{(0)}_{\ \ \ 11}=0,
\end{eqnarray}
so that $\mathcal{O}\left(c^{-2}\Omega^{2}s\right)=\mathcal{O}\left(c^{-2}\Omega^{2}s^{(0)}\right)=\mathcal{O}\left(c^{-2}\omega^{2}f^{2}\right)$ and so for the linearization of the background to be valid we require $\Omega/\omega\gg f$.
\paragraph{}
In order to study the finer aspects of the background metric, we desire to solve the corrective $\mathcal{O}\left(c^{-2}\omega\Omega f^{2}\right)$ components from \Eref{eq:kernel}, i.e. to account for the next largest terms in $\left\langle t^{(2)}_{\ \ \ ik} \right\rangle$. Preferably this may be done by further perturbing $s^{(0)}_{jk}$ to give some  $s^{(1)}_{\ \ \ jk}$ and proceeding as before, however there are two potential complications. Firstly, we must be sure the $\mathcal{O}\left(c^{-2}\omega\Omega f^{2}\right)$ terms on the LHS of \Eref{eq:kernel} are a product of a linearisable perturbation rather than those terms quadratic in the Christoffel symbols of $s^{(0)}_{\ \ \ jk}$ which we previously neglected. Secondly, we must take into account any $\mathcal{O}\left(c^{-2}\omega\Omega f^{2}\right)$ terms from the RHS of \Eref{eq:pseudotensormotivation}. The $\mathcal{O}\left(c^{-2}\omega\Omega f^{2}\right)$ relations are also to be found in \Tref{tab1}.
The neglected terms in \Eref{eq:ricci} are $\mathcal{O}\left(c^{-1}\omega^{4}\Omega^{-4}f^{4}\right)$ and so we must confine ourselves to the case $f\ll\left(\Omega/\omega\right)^{3/2}$. Turning our attention to the RHS of \Eref{eq:pseudotensormotivation}, we use the definition in \cite{vol2}:
\begin{eqnarray} \label{eq:h}
\zeta^{ikl}_{\ \ \ ,l}\propto\left[\left(-g\right)\left(g^{ik}g^{lm}-g^{il}g^{km}\right)\right]_{,m,l},
\end{eqnarray}
wishing to expand this expression in terms of the rapidly oscillating quantities of $\mathcal{O}\left(\omega\Omega f^{2}\right)$. A well-known formula for the perturbed metric determinant up to second order gives
\begin{eqnarray}
-g\approx1+s+h+\frac{1}{2}\left(s+h\right)^{2}-\frac{1}{2}\left(s^{i}_{k}+h^{i}_{k}\right)\left(s^{k}_{i}+h^{k}_{i}\right)
\end{eqnarray}  
and so any corrections to \Eref{eq:kernel} would be from
\begin{eqnarray} \label{eq:terms}
\zeta^{(2)ikl}_{\ \ \ \ \ \ ,l}=-\frac{1}{2}\left(h^{n}_{o}h^{o}_{n}\right)_{,m,l}\left(\eta^{ik}\eta^{lm}-\eta^{il}\eta^{km}\right)+\left(h^{ik}h^{lm}-h^{il}h^{km}\right)_{,m,l}.
\end{eqnarray}
The perturbation $h^{i}_{k}$ contains important terms of $\mathcal{O}\left(f\right)$ and $\mathcal{O}\left(\Omega f/\omega\right)$, so attention is confined to terms of the form $h^{a}_{b,e}h^{c}_{d,y}$ or $h^{a}_{b,y}h^{3}_{d,z}$, where $a$, $b$, $c$, $d$, $e$ are $1$ or $2$, and $y$ and $z$ are $0$ or $3$. By expanding \Eref{eq:terms} the only such terms are found not to survive averaging over a wave cycle; consequently, to this new order the RHS of \Eref{eq:pseudotensormotivation} may be neglected. One way to justify neglecting any cross terms of $s_{ik}$ and $h_{ik}$ above would be to demand $s_{ik}\ll h_{ik}$, or $f\ll \left(\Omega/\omega\right)^{2}$. The picture is clearer if we choose $h_{jk}$ to be the `fundamental' perturbation of the metric of $\mathcal{O}\left(f\right)$, in which the $s_{jk}$ is effectively second order: by always choosing sufficiently small $f$ (i.e. a weak beam) considerations of the smallness of $\Omega/\omega$ should decouple from that of $f$.
\paragraph{}
Now attention is turned to the axisymmetric beam: $f=f\left(\rho\right)$. We begin by discussing the $1$, $2$ subspace transverse to the propagation of the wave in the $s^{(0)}_{\ \ \ jk}$ approximation. It is enough to notice that the interval of a transverse, purely space-like four-vector must be isotropic about the $z$-axis to see that $s^{(0)}_{\ \ \ 11}=s^{(0)}_{\ \ \ 22}$ and $s^{(0)}_{\ \ \ 12}=0$, i.e. the background metric merely dilates the transverse space. Then \Eref{second} reduces to $\nabla^{2}_{\perp}s^{(0)}_{\ \ \ 11}=0$ and since the radially symmetric solutions to the $2$-D Laplace equation are logarithmic or constant, we apply the obvious boundary conditions on the $z$-axis and at infinity, setting $s^{(0)}_{\ \ \ 11}=0$. By analogy with \cite{light} there is expected to be some mixing of the azimuthal coordinate with both time and the axis of propagation, these corrections look in cylindrical coordinates like
\begin{eqnarray} \label{eq:perturbation2}
\left[s^{(1)}_{\ \ \ jk}\right]\bigg|_{cylindrical}= \left( 
\begin{array}{cccc}
0 & 0 & \psi\rho^2 & 0 \\
\cdot\cdot\cdot & 0 & 0 & 0 \\
\cdot\cdot\cdot & \cdot\cdot\cdot & 0 & \chi\rho^2 \\
\cdot\cdot\cdot & \cdot\cdot\cdot & \cdot\cdot\cdot & 0 \\
\end{array} 
\right),
\end{eqnarray}
or transforming to Cartesian coordinates,
\begin{eqnarray}
\left[s^{(1)}_{\ \ \ jk}\right]\bigg|_{Cartesian}= \left( 
\begin{array}{cccc}
0 & -y\psi & x\psi & 0 \\
\cdot\cdot\cdot & 0 & 0 & -y\chi \\
\cdot\cdot\cdot & \cdot\cdot\cdot & 0 & x\chi \\
\cdot\cdot\cdot & \cdot\cdot\cdot & \cdot\cdot\cdot & 0 \\
\end{array} 
\right).
\end{eqnarray}
Generally, one could posit mixing of $\rho$ and $t$, however the relevant terms disappear identically in the linearized $R_{jk}$ so it is not possible to infer them. Furthermore the wave-like perturbation that was initially enforced, $h_{jk}$, has certain symmetries which are not expected to be broken by the background metric $s_{jk}$: particularly we expect the Killing vector to be $\left(ct, 0, \varphi, z\right)$, since the beam is invariant under an inversion of time, the azimuthal coordinate and the central axis. Since however radial frame dragging would be reversed by this transformation, it may be neglected.
We begin with $\psi=\psi\left(\rho\right)$, noting that the contents of \Tref{tab1} show that $\psi$ and $\chi$ obey the same ODE:
\begin{eqnarray}
\frac{3}{\rho}\frac{\mathrm{d}}{\mathrm{d}\rho}\psi\left(\rho\right)+\frac{\mathrm{d}^{2}}{\mathrm{d}\rho^{2}}\psi\left(\rho\right)=\frac{2\omega}{c\rho}f\left(\rho\right)\frac{\mathrm{d}}{\mathrm{d}\rho}f\left(\rho\right).
\end{eqnarray}
At this point we could consider a scale\footnote{Where `sharp' edges to the transverse envelope do not contradict $\omega\gg\Omega$.} on which the beam has a uniform interior intensity\footnote{Defining the functional form of intensity rather than amplitude this way is desirable, bearing in mind the mass energy current of the toy model of a rotating dusty cylinder whose density is $\sigma\left(\rho\right)\propto\delta\left(\rho-\rho_{0}\right)$ and $\left\langle \lambda_{z}\right\rangle$ from \Eref{eq:dens}.} which sharply vanishes outside its radius, $\rho_{0}$, so that $f\left(\rho\right)^{2}=f_{0}^{2}\left(1-\Theta\left(\rho-\rho_{0}\right)\right)$, where $\Theta\left(\rho\right)$ is Heaviside's step function. This gives
\begin{eqnarray}
\psi\left(\rho\right)&=\frac{2\omega}{c}\int^{\rho}\mathrm{d}\rho'\frac{1}{\rho'^{3}}\int^{\rho'}\mathrm{d}\rho''\rho''^{2}\frac{\mathrm{d}f\left(\rho''\right)}{\mathrm{d}\rho''}f\left(\rho''\right)+\frac{c_{2}}{\rho^{2}}+c_{1}\\&
=-\frac{\omega f_{0}^{2}}{c}\int^{\rho}\mathrm{d}\rho'\frac{1}{\rho'^{3}}\int^{\rho'}\mathrm{d}\rho''\rho''^{2}\delta\left(\rho''-\rho_{0}\right)+\frac{c_{2}}{\rho^{2}}+c_{1}\\&
=-\frac{\omega f_{0}^{2}}{c}\int^{\rho}\mathrm{d}\rho'\frac{\rho_{0}^{2}}{\rho'^{3}}\Theta\left(\rho'-\rho_{0}\right)+\frac{c_{2}}{\rho^{2}}+c_{1}\\&
=\frac{\omega f_{0}^{2}\rho_{0}^{2}}{2c}\left(\frac{1}{\rho^{2}}-\frac{1}{\rho_{0}^{2}}\right)\Theta\left(\rho-\rho_{0}\right)+\frac{c_{2}}{\rho^{2}}+c_{1},
\end{eqnarray}
where $c_{1}$ and $c_{2}$ are constants. In electromagnetism, thin, long solenoids enclose uniform magnetic fields which vanish on the outside; the same can be demanded of the beam by setting $c_{2}=0$ and $c_{1}=\omega f_{0}^{2}/2c$ so that the solution is
\begin{eqnarray}
\psi\left(\rho\right)=\frac{\omega f_{0}^{2}}{2c}\left(1-\Theta\left(\rho-\rho_{0}\right)\right)+\frac{\omega f_{0}^{2}\rho_{0}^{2}}{2c\rho^{2}}\Theta\left(\rho-\rho_{0}\right).
\end{eqnarray}
Given that $\mathbf{B}_{g}=\nabla\wedge\mathbf{A}_{g}$ and $\mathbf{A}_{g}/c=\left(-y\psi\left(\rho\right),x\psi\left(\rho\right),0\right)$ we have $\mathbf{B}_{g}=\omega f_{0}^{2}\mathbf{\hat{z}}$ within and zero without.
\begin{figure}[t] \label{fig2}
\centering
\includegraphics[width=12cm]{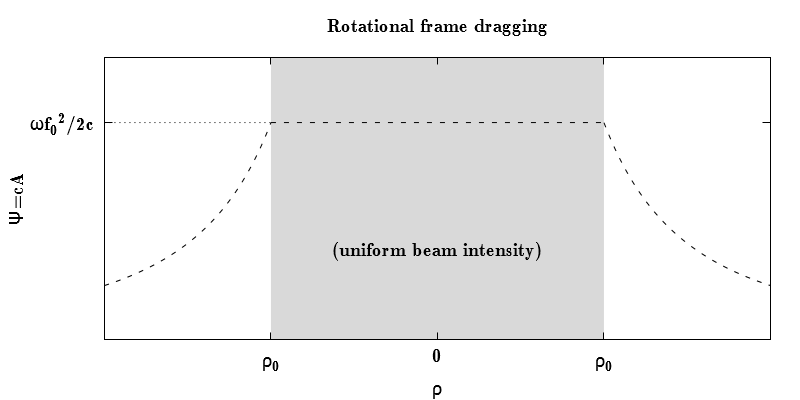}
\caption{A longitudinal section of the sharply-defined beam, showing the rate of rotational frame-dragging.}
\label{fig:rates}
\end{figure}
\paragraph{}
By integrating the angular momentum density over the transverse plane and recalling the choice of $f\left(\rho\right)$ we see that the density of linear momentum in the `skin' of the beam must be $\left| \textbf{p}\right| =c^{2}\omega f_{0}^{2}/16\pi G$ and applying the gravomagnetic Biot Savart law for a presumably uniform axial $B_{g}$ field gives the same result immediately. The beam is therefore understood to be a gravomagnetic solenoid. Neglecting $\mathcal{O}\left(\psi^{2}\right)$ time-dilation effects as extra small, the mixed space-time corrections in \Eref{eq:perturbation2} represent a rotation of the inerial frame at angular velocity, $\Psi=c\psi\left(\rho\right)$. This may be compared with the formula $\Psi=4Gl_{z}/c^{2}\rho_{0}^{2}$ from \cite{cylinders} which describes the same effect for a general mass-energy cylinder of angular momentum per unit length, $l_{z}=\int\int\lambda_{z}\mathrm{d}x\mathrm{d}y$. These effects are illustrated in \Fref{fig2}.
\par
That this is not a \textit{gravomagnetic} effect outside the beam is because the rotational dragging of the frame about the beam cancels exactly with the shearing of the frames, the tidal torques, at every point; contrast this with Kerr spacetime where the principal gravomagnetic effect is not that of frame dragging around the black hole, but of such tidal torques, which act in a counter-rotating sense to the source. There are no tidal torques within the beam, and the uniform rotation of the inertial frame may therein be cast as a gravomagnetic field.
\par
For $\chi=\chi\left(\rho\right)$, taking $c_{1}=c_{2}=0$ gives
\begin{eqnarray}
\chi\left(\rho\right)=\frac{\omega f_{0}^{2}\rho_{0}^{2}}{2c}\left(\frac{1}{\rho^{2}}-\frac{1}{\rho_{0}^{2}}\right)\Theta\left(\rho-\rho_{0}\right).
\end{eqnarray}
This choice of integration constants is indicated by \cite{cylinders2}, where the metric external to the thin material cylindrical shell carrying a torque as well as linear momentum is shown to be twisted into a helix whose pitch (or $\chi$) vanishes at $\rho_{0}$ to match the interior and approaches a constant at infinite radius.
\newpage

\ack
\par
I am grateful to support of Grant LGOH>EFKM, the hospitality of the Institute of Astronomy, Cambridge and the vital discussions with D Lynden-Bell and J Bi\v{c}\'{a}k.
\clearpage
\section*{References}

\maketitle
\end{document}